\documentclass{ws-procs975x65}
\def\be{\begin{equation}}
\def\ee{\end{equation}}
\def\dd{\mathrm{d}}
\def\rr{\mathrm}
\def\mpl{m_\rr{pl}}

\def\rhoA{\rho_{\rr {A}}}
\begin{document}

\title{Numerical forecasts for lab experiments constraining modified gravity: the chameleon model}

\author{Sandrine Schl\"ogel$^*$}
\address{naXys, University of Namur,\\ Rempart de la Vierge 8,
Namur, 5000, Belgium\\
$^*$E-mail: sandrine.schlogel@unamur.be\\
www.unamur.be}
\address{Centre for Cosmology, Particle Physics and Phenomenology,
Institute of Mathematics and Physics, Louvain University,
Chemin du Cyclotron 2, 1348 Louvain-la-Neuve, Belgium}

\author{S\'ebastien Clesse}
\address{naXys, University of Namur,\\ Rempart de la Vierge 8,
Namur, 5000, Belgium}
\address{Institute for Theoretical Particle Physics and Cosmology (TTK), RWTH Aachen University, D-52056 Aachen, 
Germany}

\author{Andr\'e F\"uzfa}
\address{naXys, University of Namur,\\ Rempart de la Vierge 8,
Namur, 5000, Belgium\\
$^*$E-mail: sandrine.schlogel@unamur.be\\
www.unamur.be}
\address{Centre for Cosmology, Particle Physics and Phenomenology,
Institute of Mathematics and Physics, Louvain University,
Chemin du Cyclotron 2, 1348 Louvain-la-Neuve, Belgium}

\begin{abstract}
Current acceleration of the cosmic expansion leads to coincidence as well as fine-tuning 
issues in the framework of general relativity. Dynamical scalar fields 
have been introduced in response of these problems, some of them invoking screening mechanisms for passing 
local tests of gravity. Recent lab experiments based on atom interferometry in a vacuum chamber have 
been proposed for testing modified gravity models.  So far only analytical computations have been used to 
provide forecasts. We derive numerical solutions for chameleon models that take into account the effect of the 
vacuum chamber wall and its environment. With this realistic profile of the chameleon field in the 
chamber, we refine the forecasts that were derived analytically. We finally highlight specific 
effects due to the vacuum chamber that are potentially interesting for future experiments.
\end{abstract}


\bodymatter

%%%%%%%%%%%%%%%%% now a standard article style for the most part
\section{Introduction}
A challenging issue in cosmology today consists in explaining the current accelerated cosmic expansion or 
'dark energy'. Even if the standard model of cosmology reproduces current observations, the cosmological 
constant can not explain the coincidence issue and faces a fine-tuning problem. The most 
simple 
alternative is to introduce a dynamical scalar field, potentially originating from the gravity sector. 
However, such 
models are a dangerous business since they have to pass stringent constraints in the Solar 
system and in lab experiments. To do so, models invoking a 
screening mechanism, like the chameleon\cite{KhouryWeltmanPRL,KhouryWeltmanPRD}, have been built: in dense 
environment like the Solar system, the scalar field is suppressed while it acts on sparse environment, like 
in the cosmos at late time. 

Recently, a new lab experiment based on atom interferometry with a test mass inside a vacuum chamber has 
been designed for testing modified gravity models\cite{KhouryHamilton}. The idea is that, even if their 
nuclei appear to be dense, atoms are so small that the scalar field is unsuppressed. Additional acceleration 
on 
individual atoms due to the chameleon field gradient inside the vacuum chamber in the presence of the test 
mass could be measured. Analytical calculations derived so far in Refs.~\refcite{KhouryHamilton},~\refcite{Burrage} 
show 
that most
of the chameleon parameter space is constrained by such an experiment. 

Nevertheless, the authors assumed that the wall effect is negligible and the value of the scalar field at the center of 
the chamber is 
then determined as a function of the size of the chamber\cite{Burrage,KhouryHamilton}. We 
provide numerical computations with the following minimal assumption\cite{Nous}: the chameleon field reaches its 
equilibrium 
value $\phi_\infty$ in the outside atmosphere. Our results reveal that the scalar field 
amplitude inside the chamber is related to $\phi_\infty$ instead of the chamber size, in the case where the 
test mass perturbs 
weakly the chameleon profile. We also study the \textit{strongly} perturbing case where \textit{thin shell} 
appears. In both cases, we provide forecasts for the acceleration due to the scalar field $a_\phi$  which is related to 
the scalar field gradient inside the vacuum chamber, experimental constraint\cite{KhouryHamilton} being $a_\phi/g<5.5\times10^{-7}$, 
with $g$ the Earth gravitational acceleration.
We also highlight the effects of the test mass density and size, a result which can be interesting for 
designing further experiments.

\section{The chameleon model}
In this section, we briefly remind the chameleon model. We start from the general scalar-tensor theory 
action written in the Einstein frame,
\be
S=\int \dd^4x \sqrt{-g} \left[\frac{R}{2\kappa}-\frac{1}{2} \left(\partial\phi\right)^2-V(\phi)\right]
+ S_m\left[A^2\left(\phi\right)g_{\mu\nu};\psi_\rr{m}\right],
\ee
with $R$, the scalar curvature, $\kappa=8\pi/\mpl^2$, $\mpl$ being the Planck mass, $\psi_\mathrm{m}$ the 
matter fields, $V(\phi)$ the chameleon potential (in the following, we will consider 
$V(\phi)=\Lambda^{\alpha+4}/\phi^\alpha$, 
$\alpha$ and $\Lambda$ being the parameters of the potential) and $A(\phi)$ a general coupling function (in 
the following, we will consider $A(\phi)=\rr{e}^{\phi/M}$, $M$ being a parameter). For a static 
and a spherically symmetric spacetime in the non-relativistic limit, the Klein-Gordon equation is given by,
\be 
\phi''+\frac{2}{r} \phi'={\frac{\dd V_{\rm eff}}{\dd \phi}}, \hspace{1cm} {\frac{\dd V_{\rm 
eff}}{\dd \phi}}={\frac{\dd V}{\dd \phi}} + 
\tilde{\rho} A^3 {\frac{\dd A}{\dd \phi}},
\ee
with $V_{\rm eff}$ the effective potential, $\tilde\rho$ the density written in the Jordan frame\cite{Hees} such as it 
obeys to the energy conservation $\nabla_\mu \tilde{T}^{\mu\nu}=0$, a prime denoting a derivative 
with 
respect to the radial coordinate. The effective potential minimum $\phi_{\rr{min}}$ is given by,
\be
\phi_{\rr{min}}=\left(\frac{\alpha \Lambda^{\alpha+4}M}{\tilde{\rho}}\right)^{1/(\alpha+1)},
\ee
in the limit $A(\phi)\approx1$. Depending on the environment, the scalar field is suppressed 
(dense environment) or not (sparse environment).
\section{Numerical results}
In order to constrain the acceleration due to the scalar field $a_\phi=\partial_r\phi/M$, we solve the 
Klein-Gordon equation inside the test mass, the vacuum chamber, the wall and in the air outside the vacuum 
chamber with the minimal assumptions: the scalar field reaches its equilibrium value 
$\phi_\infty=\phi_\rr{min}\left(\tilde{\rho}_\rr{air}\right)$ at spatial infinity and the solution should 
be regular at the origin of coordinates and everywhere continuous. We use a solver for multi-point boundary 
value problem with unknown parameters\cite{Matlab}, the boundary conditions being given by $\phi'(r=0)=0$ and 
a Yukawa profile far in the exterior environment,
\be
\phi=\phi_{\infty}+\frac{\mathcal{C}{\rm e}^{-\mathcal{M}r}}{r},
\ee
with $\mathcal{M}=\left.\dd^2V_\rr{eff}/\dd\phi^2\right|_{\phi=\phi_\infty}$, the constant 
of integration $\mathcal{C}$ being a parameter to be determined by the numerical algorithm.
\subsection{Acceleration forecasts}
\begin{figure}
\begin{center}
\includegraphics[scale=0.3, trim= 50 0 100 0, clip=true]{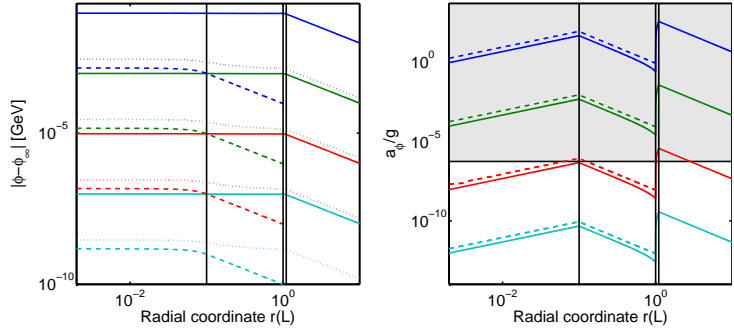}
\caption{Scalar field $|\phi-\phi_\infty|$ and acceleration $a_\phi$ profiles ($\alpha=1$ and $\Lambda=2.6 \times 
10^{-6}$~GeV) for various $M$ ($M=10^{13},10^{15},10^{17},10^{19}$ GeV in blue, green, red and light blue). Solid and 
dashed lines refer to numerical (4 regions) and analytical (2 regions) profiles respectively while dotted lines are 
obtained when lowering the wall density by a factor $10^2$. The vertical lines 
mark out the four regions (test mass, vacuum chamber, chamber wall and outside)\cite{Nous}.}
\centering
\label{plot:profiles}
\end{center}
\end{figure}

\begin{figure}
\begin{center}
\includegraphics[scale=0.25, trim= 250 0 300 0, clip=true]{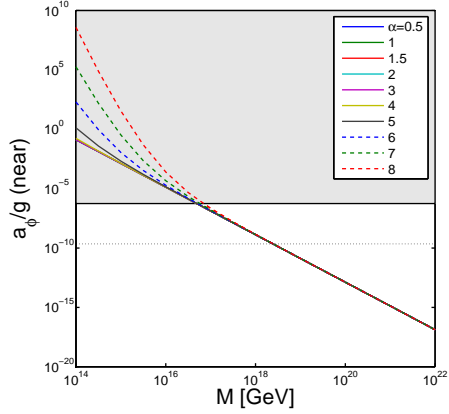}
\caption{Forecast for the normalized acceleration $a_\phi/g$ measured in the \textit{near} position, i.e. 
$8.8$ mm far from the test mass, for various $M$ and $\alpha$. Dotted line represents the acceleration due to 
the gravitational interaction with the test mass. The gray rectangle shows which part of the parameter space 
is ruled out \cite{Nous}.}
\centering
\label{plot:simu_near}
\end{center}
\end{figure}

We compare the numerical and analytical profiles of the scalar field and the acceleration on Fig.~\ref{plot:profiles} 
for various $M$ ($\alpha=1$ and corresponding $\Lambda$ being obtained from the cosmological 
constraints coming from SNe Ia observations\cite{Hees}). The numerical results differ by up to one order of magnitude 
for the acceleration compared to previous analysis\cite{Burrage, KhouryHamilton}, indicating that the effect of the wall 
cannot be neglected in a precise investigation of the chameleon parameter space. Indeed, varying the wall density by two 
orders of magnitude (see dotted lines on the scalar field profile of Fig.~\ref{plot:profiles}), we show that the wall 
perturbs more or less importantly the field profile. The effect remains however negligible for the acceleration profile 
that is related to the gradient of the field. 
Another important result is the determination of the central value of the scalar field: in Ref.~\refcite{Burrage}, it 
is given by the size of the chamber while the numerical simulations show that it is better approximated by 
$\phi_\infty=\phi_\rr{min}\left(\tilde{\rho}_\rr{air}\right)$.
On Fig.~\ref{plot:simu_near}, we study the effect of 
$\alpha$ and $M$ on the acceleration at 8.8 mm far from the test mass (refers 
as the \textit{near} position in Ref.~\refcite{KhouryHamilton}) where the acceleration is measured experimentally.  We 
conclude that the experiment presented in Ref.~\refcite{KhouryHamilton} is able to rule out the chameleon model 
presented 
in the previous section for $M\lesssim 10^{17}$ GeV whatever $\alpha$. Discrepancies due to $\alpha$ in the 
determination of the acceleration highlighted on Fig.~\ref{plot:simu_near}, appears when the limit $|A(\phi)-1|\ll 1$ is 
no more valid. 
We also studied the thin shell regime and our results validate the analytical approximations to a good accuracy.

\subsection{Geometry effects}
\begin{figure}
\begin{center}
\includegraphics[scale=0.3, trim= 200 0 250 0, clip=true]{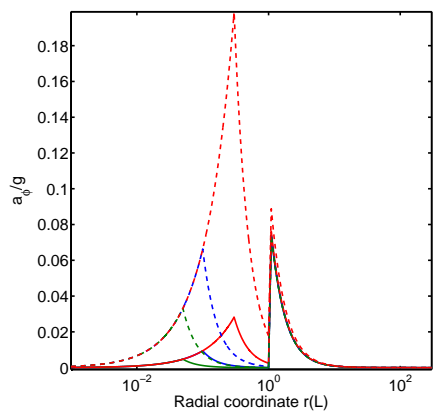}
\caption{Comparison of chameleon acceleration $a_\phi/g$ depending on which material the test mass is made of.
  Solid and dashed lines correspond to aluminum and tungsten $\rhoA$ respectively while green, blue and red colors 
 correspond to 5 mm, 1 cm and 3 cm test mass radius. We fix here $\alpha=1$ and $M=10^{15}$ GeV.}
\centering
\label{plot:geom}
\end{center}
\end{figure}

As stated in the Sec.~I, in the case where the scalar field is \textit{weakly} perturbed by the test mass, we 
observe only a small deviation with respect to $\phi_\infty$. This is due to the presence of the wall 
chamber. It stabilizes the scalar field and gives it a kick for reaching $\phi_\infty$. 
Furthermore, on Fig.~\ref{plot:profiles}, we see that the wall density is responsible for a variation of two 
orders of magnitude in the scalar field profiles. Since in the experimental setup proposed in 
Ref.~\refcite{KhouryHamilton} the wall and test mass are similar in size and density, we expect similar effects while 
varying test mass density and size. The acceleration profiles for the test mass made of aluminum and tungsten with a 
radius of 5 mm, 1 cm and 3 cm
are reported on Fig.~\ref{plot:geom}. We show that, choosing a test mass which is denser and bigger, the acceleration can differ by almost a factor 10.

\section{Conclusion}
We derived numerically forecasts for the experiment of Ref.~\refcite{KhouryHamilton} for various chameleon 
models in the \textit{weakly perturbing regime}. We showed that analytical and numerical acceleration forecasts and 
constraints differ by up to one order of magnitude. We also highlight that the numerical simulations can be helpful for 
precise investigation of the parameter space of the chameleon models as well as for optimizing the experimental setup. 
The same numerical method has been used to derive constraints on other chameleon potentials in Ref.~\refcite{Nous} for 
the thin shell regime. Our numerical method could be easily extended to other modified gravity models 
like the symmetron.

\section*{Acknowledgments}
We warmly thank Holger M\"uller for the discussion during the MG meeting and the following conversation where Justin 
Khoury, Benjamin Elder and Philipp Haslinger took part. We also warmly thank Clare Burrage and Christophe Ringeval   
for useful comments and discussion. S.S. is supported by the FNRS-FRIA, S.C. is partially supported by the 
\textit{Return Grant} program of the Belgian Science Policy (BELSPO) and A. F. is partially supported by the ARC 
convention No. 11/15-040.

\end{document}